\documentclass[conference]{IEEEtran}

  	\usepackage[pdftex]{graphicx}
  	\graphicspath{{../pdf/}{../jpeg/}}
	\DeclareGraphicsExtensions{.pdf,.jpeg,.png}

	\usepackage[cmex10]{amsmath}
	\usepackage{mathabx}
	\usepackage{algorithmic}
	\usepackage{array}
	\usepackage{mdwmath}
	\usepackage{mdwtab}
	\usepackage{eqparbox}
	\usepackage{url}
        \usepackage{chngcntr}
        \usepackage{graphicx}
        \usepackage{subcaption}

\begin{document}

\title{\LARGE The impact of parameter spread of high-temperature superconducting Josephson junctions on the performance of quantum-based voltage standards}

 \author{\authorblockN{Guanghong Wen, Yi Zhu, Yingxiang Zheng, Shuhe Cui, Ji Wang, Yanyun Ren,\\Hao Li*, Guofeng Zhang, Lixing You }
 \authorblockA{Shanghai Institute of Microsystem and Information Technology, CAS, Shanghai 200050\\haoli@mail.sim.ac.cn}}
    
\maketitle

\begin{abstract}
Quantum metrology based on Josephson junction array reproduces the most accurate desired voltage by far, therefore being introduced to provide voltage standards worldwide. In this work, we quantitatively analyzed the dependence of the first Shapiro step height of the junction array at 50 GHz on the parameter spread of 10,000 Josephson junctions by numerical simulation with resistively shunted junction model. The results indicate an upper limit spread of the critical current and resistance of the Josephson junctions. Specifically, to keep the maximum first Shapiro step above 0.88 mA, the critical current standard deviation, $\sigma$, should not exceed 25\%, and for it to stay above 0.6 mA, the resistance standard deviation should not exceed 1.5\%.

\end{abstract}

\IEEEoverridecommandlockouts
\begin{keywords}
Josephson junction array; Voltage standard; High-temperature superconducting; Parameter spread
\end{keywords}

\IEEEpeerreviewmaketitle


\section{Introduction}
The discovery of the Josephson effect \cite{josephson1962possible} and the subsequent development of the Josephson voltage standard (JVS) \cite{taylor1967use}\cite{1971ac}\cite{1973Volt} have significantly improved the precision of voltage references. Initially, single-junction JVS produced relatively low voltages, typically below 10 mV. To achieve higher voltages, multiple junctions are connected in series. In 1983, Endo et al.\cite{Endo1983High} demonstrated a series array of 20 junctions, employing individual bias currents to address parameter inconsistencies, resulting in an output of 100 mV with an uncertainty of $10^{-9}$. However, scaling this method to larger arrays is impractical, and the key challenge lies in improving the parameter consistency of the Josephson junctions themselves.

Advancements in novel fabrication techniques and micro-nano processing have significantly improved the parameter consistency of low-temperature superconducting (LTS) Josephson junctions\cite{1995Superconductor}\cite{Schulze1998Nb}\cite{2001All}, facilitating the development of programmable arrays capable of achieving output voltages of up to 10 V\cite{1997Stable}\cite{200610}\cite{Paul201010}. However, the precision measurement and metrology industries have not been revolutionized by the LTS-based JVS because of its high-cost and non-portable features originated from the high-power and heavy cryocooler. Consequently, the development of a high-temperature superconducting (HTS) Josephson junction array voltage standard that can operate within the liquid nitrogen temperature range has become of significant importance.

Early fabrication methods for HTS Josephson junctions include bicrystal\cite{1988Orientation}, step-edge\cite{1991Engineered,1991Practical}, and ramp-edge\cite{1990Controlled} techniques, which laid the foundation for initial developments in the field. More recently, advancements in Helium-Focused Ion Beam (He-FIB) technology\cite{2015Nano}\cite{2018Direct}\cite{2024Arrays} have enabled the precise and efficient fabrication of HTS Josephson junction arrays, creating favorable conditions for the realization of HTS-based voltage standards. However, achieving the high parameter uniformity required for practical applications remains a challenge, as experimental studies show that the uniformity of current HTS Josephson junctions still falls short of that achieved by their LTS counterparts\cite{1999High}\cite{2003Evaluation}\cite{2014Critical}. To address this challenge, we employ numerical simulation methods to establish a practical benchmark for the parameter uniformity required for the implementation of HTS-based JVS, thereby providing a clear target for future research and guiding efforts toward achieving the desired parameter consistency in
HTS Josephson junctions.


\section{Simulation model and method}
The principle behind using Josephson junctions in voltage standards is the Josephson effect, as described by the Josephson equation: 

\begin{equation}
\begin{cases}I=I_c\sin\left(\dfrac{2e}{\hbar}\int Vdt\right)\\\dfrac{\partial\varphi}{\partial t}=\dfrac{2e}{\hbar}V\end{cases}
\label{eq2.1}
\end{equation}

\begin{figure}[ht!] 
\centering
\includegraphics[width=2.5 in]{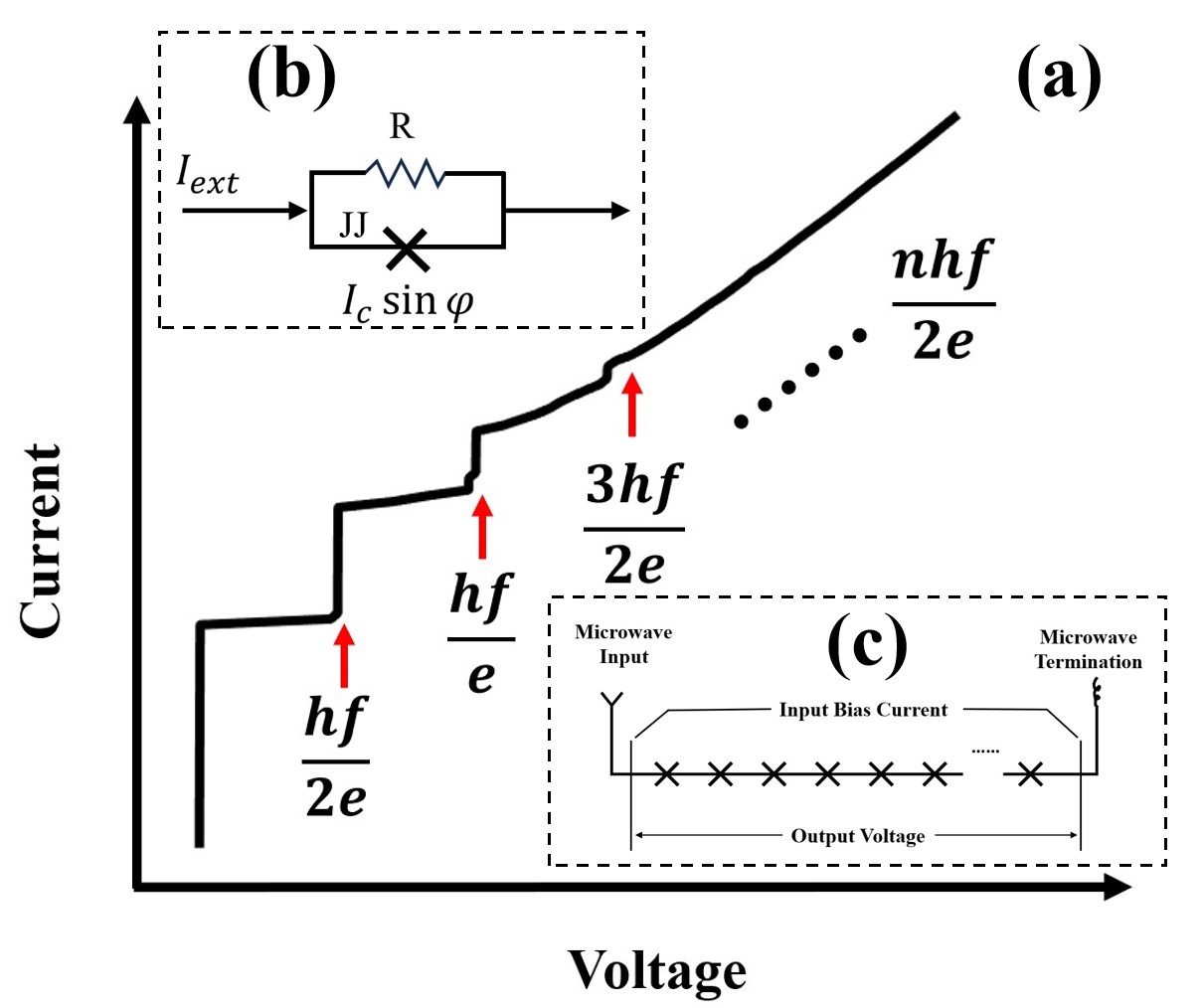}
\caption{(a) Shapiro steps observed in the $I-V$ characteristic of the Josephson junction. (b) Schematic representation of the RSJ model. (c) Schematic diagram of a Josephson junction array\cite{hamilton1997josephson}.}
\label{fig1}
\end{figure}

In this equation, $I$ represents the junction current, $I_c$ is the critical current, $V$ denotes the junction voltage, $e$ is the elementary charge, $\hbar$ is the reduced Planck constant and $\varphi$ is the phase difference between the macroscopic wave functions of the Cooper pairs in the two superconductors. Equation \eqref{eq2.1} indicates that when a dc voltage is applied across the junction, the junction current will oscillate at a frequency of $f_{J}=2eV/h$, where $ 2 e / h \approx 484 \mathrm{GHz} / \mathrm{mV}$. Due to the extremely high oscillation frequency, such oscillations are typically difficult to observe directly. However, if an ac current with frequency $f$ is applied to the junction, the oscillation will phase lock to this external frequency, resulting in a constant voltage step appearing at $V = hf/2e$ on the current-voltage ($I-V$) curve of the junction. Furthermore, the junction can synchronize with harmonics of $f$, leading to a series of steps at $V_{n} = nhf/2e$, where $n$ is an integer. These additional steps, shown in Figure \ref{fig1}(a), are commonly referred to as Shapiro steps\cite{shapiro1963josephson}.

The relationship between current and voltage in a Josephson junction can be described using the resistively shunted junction (RSJ) model\cite{stewart1968current}\cite{mccumber1968effect}, as shown in the schematic diagram in Figure \ref{fig1}(b). According to Kirchhoff's current law (KCL) and in combination with Equation \eqref{eq2.1}, the current-voltage relationship of the RSJ model can be derived, as shown in Equation \eqref{eq2.2}. When microwave irradiation with frequency $f$ and current amplitude $I_{rf}$ is applied to the junction, Equation \eqref{eq2.2} can be further expressed as Equation \eqref{eq2.3}.

\begin{equation}
  I_{ext}(t)=I_c\sin\varphi(t)+\frac{V(t)}{R}
  \label{eq2.2}
\end{equation}

\begin{equation}
  I_{ext}(t)+I_{rf}\sin(2\pi ft)=I_c\sin\varphi(t)+\frac{V(t)}{R}
  \label{eq2.3}
\end{equation}

Due to the small voltage generated by a single Josephson junction, multiple junctions must be connected in series, as shown in Figure \ref{fig1}(c). This configuration increases the output voltage, thus improving the accuracy of the voltage standard. It is important to note that each of these Josephson junctions is modeled using the RSJ framework.

\begin{figure}[ht!]
  \centering
  \begin{subfigure}{0.24\textwidth}
    \centering
    \includegraphics[width=\linewidth]{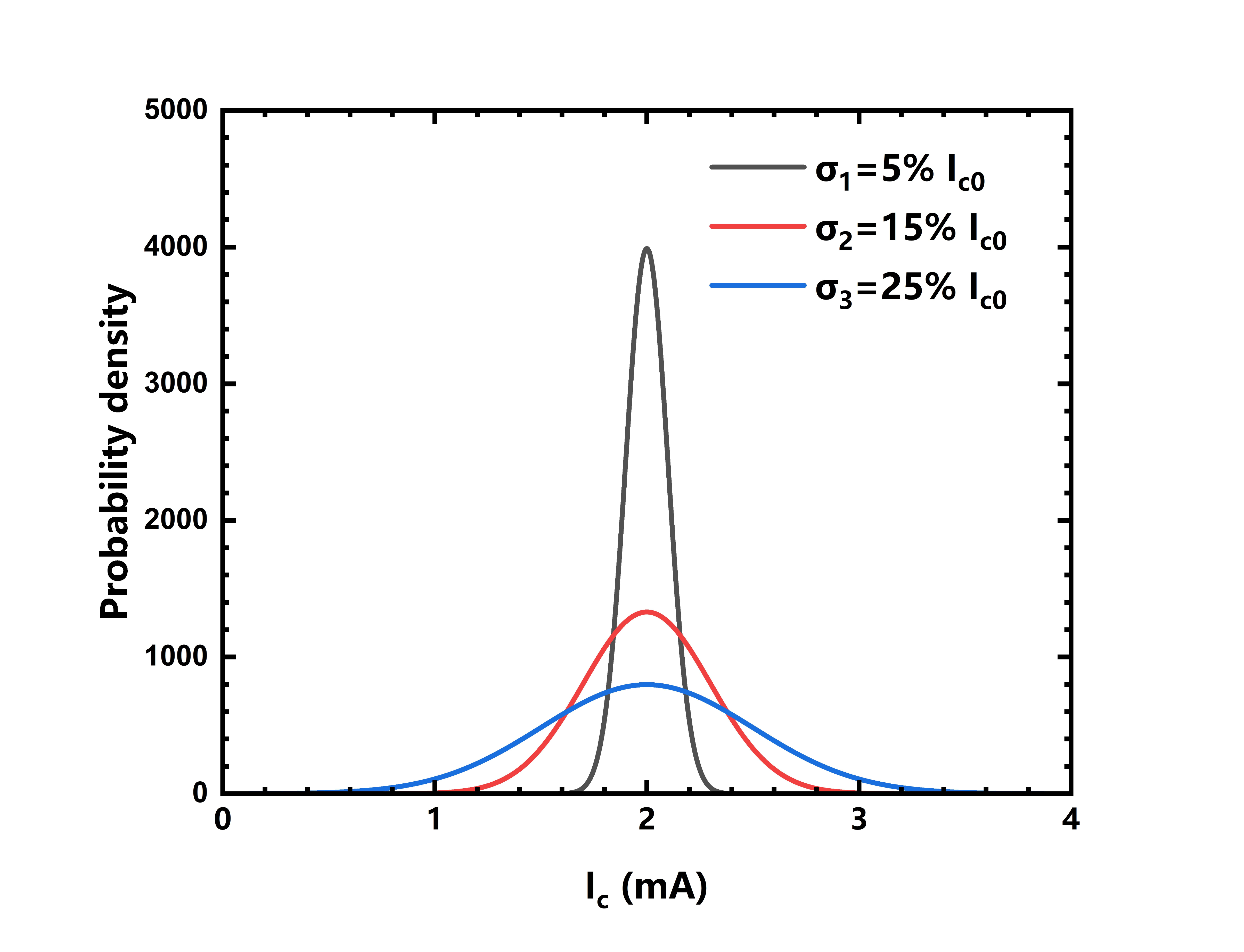}
    \caption{}
    \label{fig2a}
  \end{subfigure}\hfill
  \begin{subfigure}{0.24\textwidth}
    \centering
    \includegraphics[width=\linewidth]{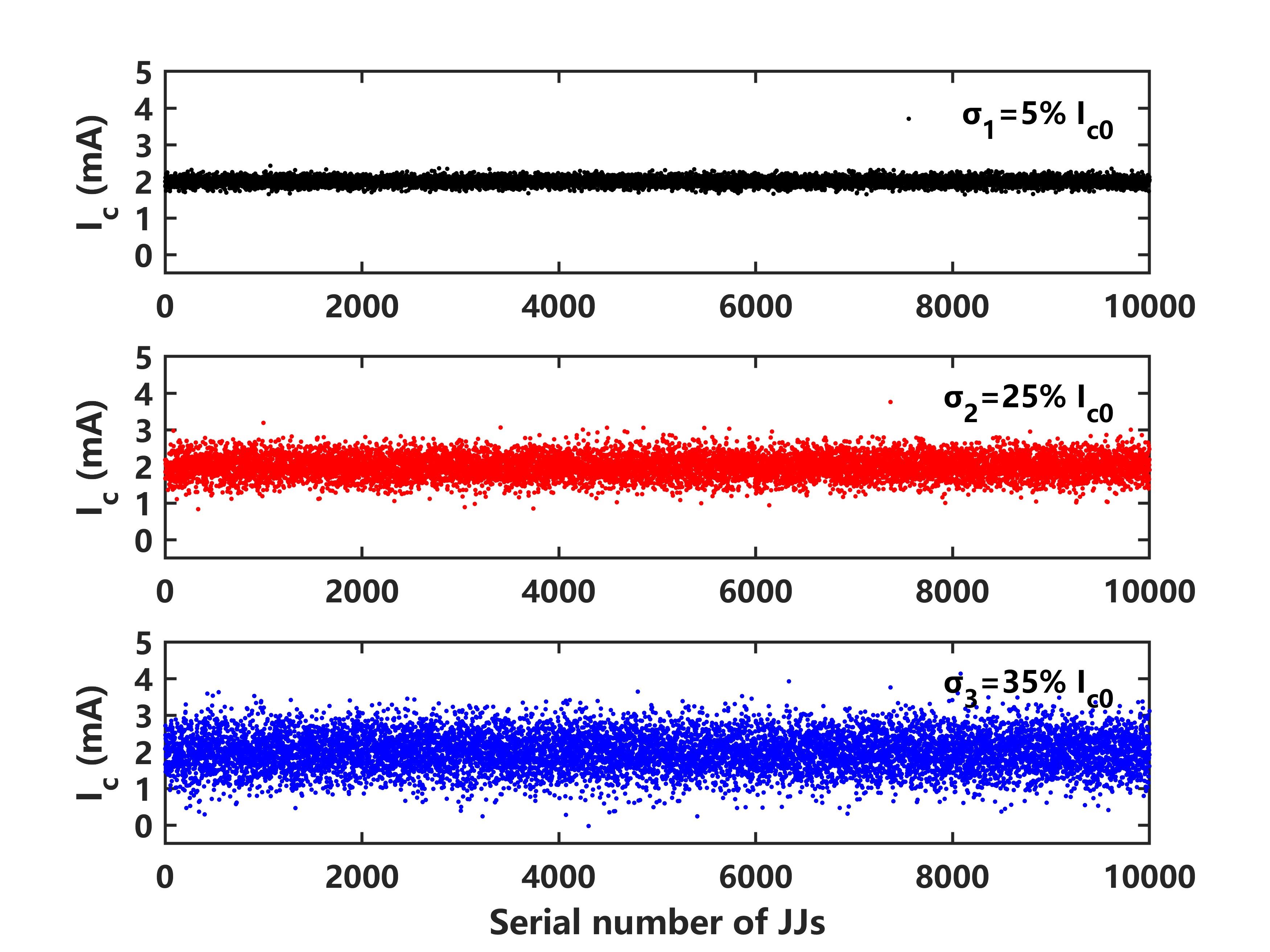}
    \caption{}
    \label{fig2b}
  \end{subfigure}
  \caption{The critical current spread of the Josephson junctions follows a normal distribution, with (a) representing the probability density curve and (b) depicting the scatter plot under different standard deviations $\sigma$.}
  \label{fig2}
\end{figure}

Several researchers have conducted experimental studies to investigate the distribution of critical current $I_c$ in Josephson junction arrays, including both LTS and HTS junctions fabricated using various processes\cite{2014Critical}\cite{1999High}\cite{2003Evaluation}\cite{kasaei2019reduced}. The experimental results indicate that, under conditions involving a large number of junctions, the spread of critical currents follows a Gaussian (normal) distribution. Based on this, the present study assumes a normal distribution for the critical current in the Josephson junction array, with $\mu = I_{c0}$ representing the average critical current. By varying the standard deviation $\sigma$, the impact of different critical current distributions on step heights is investigated. The probability density curve of the normal distribution and the scatter plot of critical currents are shown in Figure \ref{fig2}.

In this study, we numerically solved Equation \eqref{eq2.3} using the Euler method based on difference equations in MATLAB. A time step of $\Delta t = 1 \times 10^{-12}  \text{ s}$ was employed, with the total simulation time set to $t = 1 \times 10^{-7}  \text{ s}$, resulting in 100,001 time steps. Specifically, we investigated an array consisting of 10,000 junctions connected in series. The average critical current was set to $I_{c0} = 2  \text{ mA}$, with different standard deviations $\sigma$ of 5\%, 15\%, 25\% and 35\% of $I_{c0}$. The microwave frequency was fixed at 50 GHz, yielding a first Shapiro step voltage of $V_1 = 1.034 \text{ V}$ according to $V_{n} = nhf/2e$. To achieve impedance matching with other microwave components, the total resistance of the array was set to 50 $\Omega$. It should be noted that verification results demonstrate the simulation accuracy is closely related to the total simulation time $t$. While longer simulation time yield higher accuracy, they simultaneously demand exponentially increasing computational resources. Therefore, after comprehensive consideration of both accuracy and computational requirements, this study adopts the aforementioned simulation parameters for numerical simulations. All computational simulations were performed on a high-performance computing platform equipped with 30-core X86-architecture processors operating at 2.5 GHz.

\begin{figure}[ht!] 
\centering
\includegraphics[width=2 in]{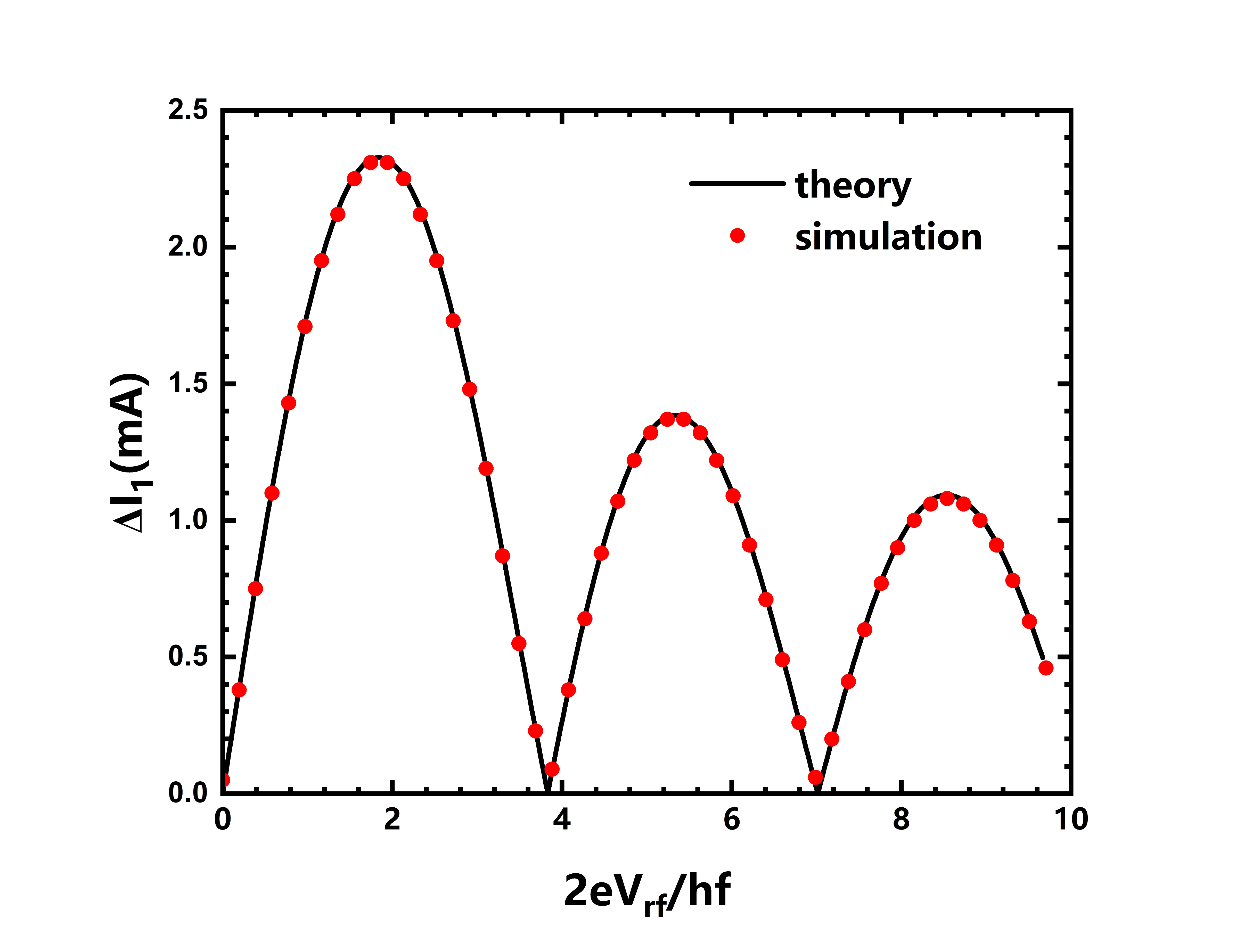}
\caption{Comparison of the theoretical values (black solid line) and the numerical simulation results (red dots).}
\label{fig3}
\end{figure}

Theoretically, the height of the first Shapiro step, $\Delta I_1$, is determined by the critical current $I_c$, the microwave voltage amplitude $V_{rf} = R I_{rf}$, and the microwave frequency $f$. The specific relationship is given by Equation \eqref{eq2.4}, where $J_1$ represents the first-order Bessel function of the first kind\cite{2009Application}. To validate the accuracy of the numerical computations, we compared the simulated values of $\Delta I_1$ with the theoretical predictions. As shown in Figure \ref{fig3}, the numerical results closely match the theoretical values, confirming the correctness of the computational methodology and program code used.

\begin{equation}
  \Delta I_1=2I_c\left|J_1\left(\frac{2eV_{rf}}{hf}\right)\right|
  \label{eq2.4}
\end{equation}

\section{Results and Discussion}

With the inclusion of thermal noise, Equation \eqref{eq2.3} is modified to Equation \eqref{eq3.1}, where $I_N(t)$ represents the Nyquist noise.

\begin{equation}
  I_{ext}(t)+I_{rf}\sin(2\pi ft)+I_N(t)=I_c\sin\varphi(t)+\frac{V(t)}R
  \label{eq3.1}
\end{equation}

It is well-established that thermal noise current lacks temporal correlation, making it a frequency-independent noise (referred to as white noise) with a mean value of zero and an amplitude that follows a normal distribution. In the subsequent simulations, the noise is modeled as a discrete random sequence with a mean square deviation $\langle \Delta I_{N}^2 \rangle = 2 k_B T/R \Delta t$ and a mean value $\langle I_N \rangle = 0$, where $T = 77 \text{ K}$ and $k_B$ is the Boltzmann constant\cite{granata2016nano}.

\begin{figure}[ht!]
  \centering
  \begin{subfigure}{0.16\textwidth}
    \centering
    \includegraphics[width=\linewidth]{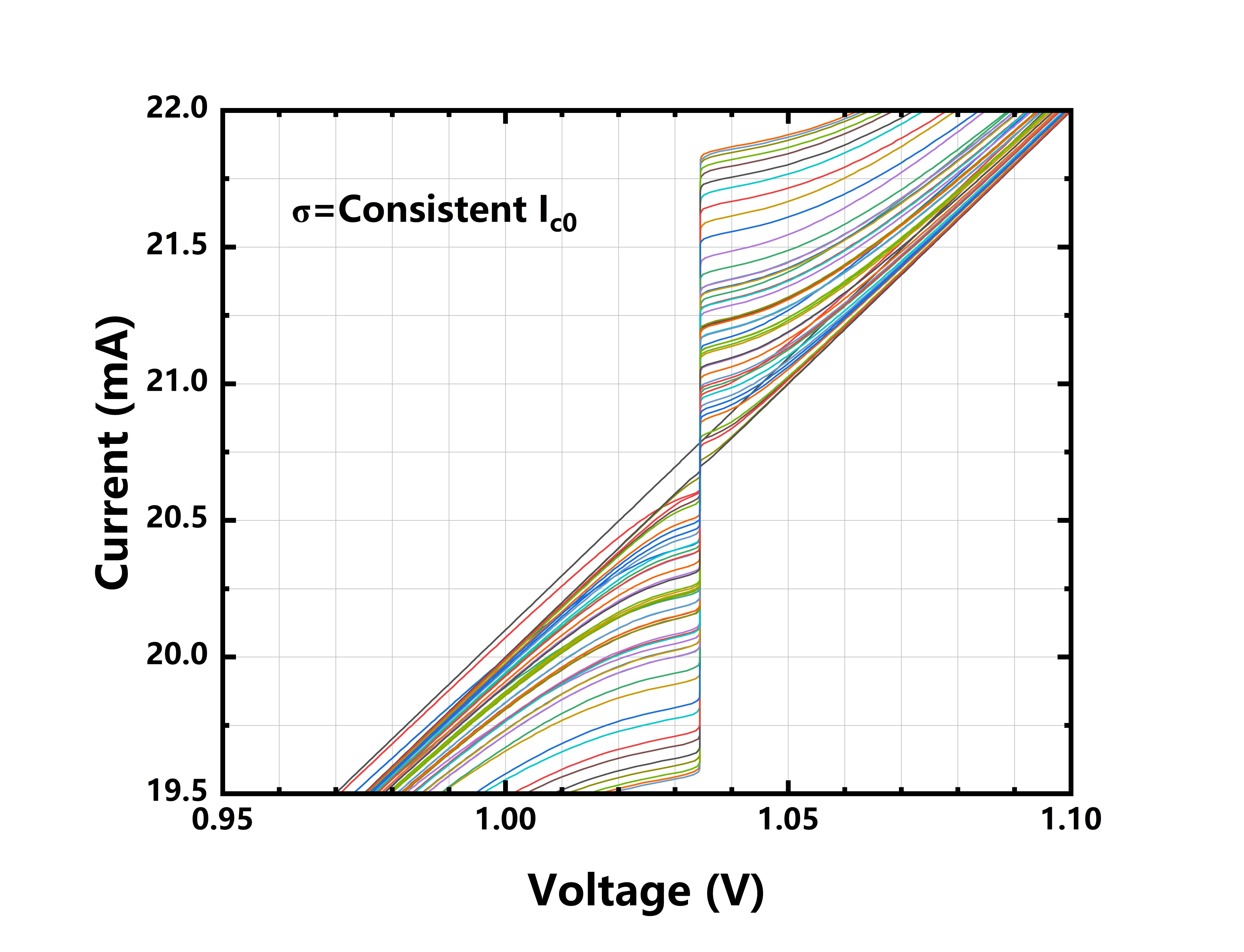}
    \caption{}
    \label{fig4a}
  \end{subfigure}\hfill
  \begin{subfigure}{0.16\textwidth}
    \centering
    \includegraphics[width=\linewidth]{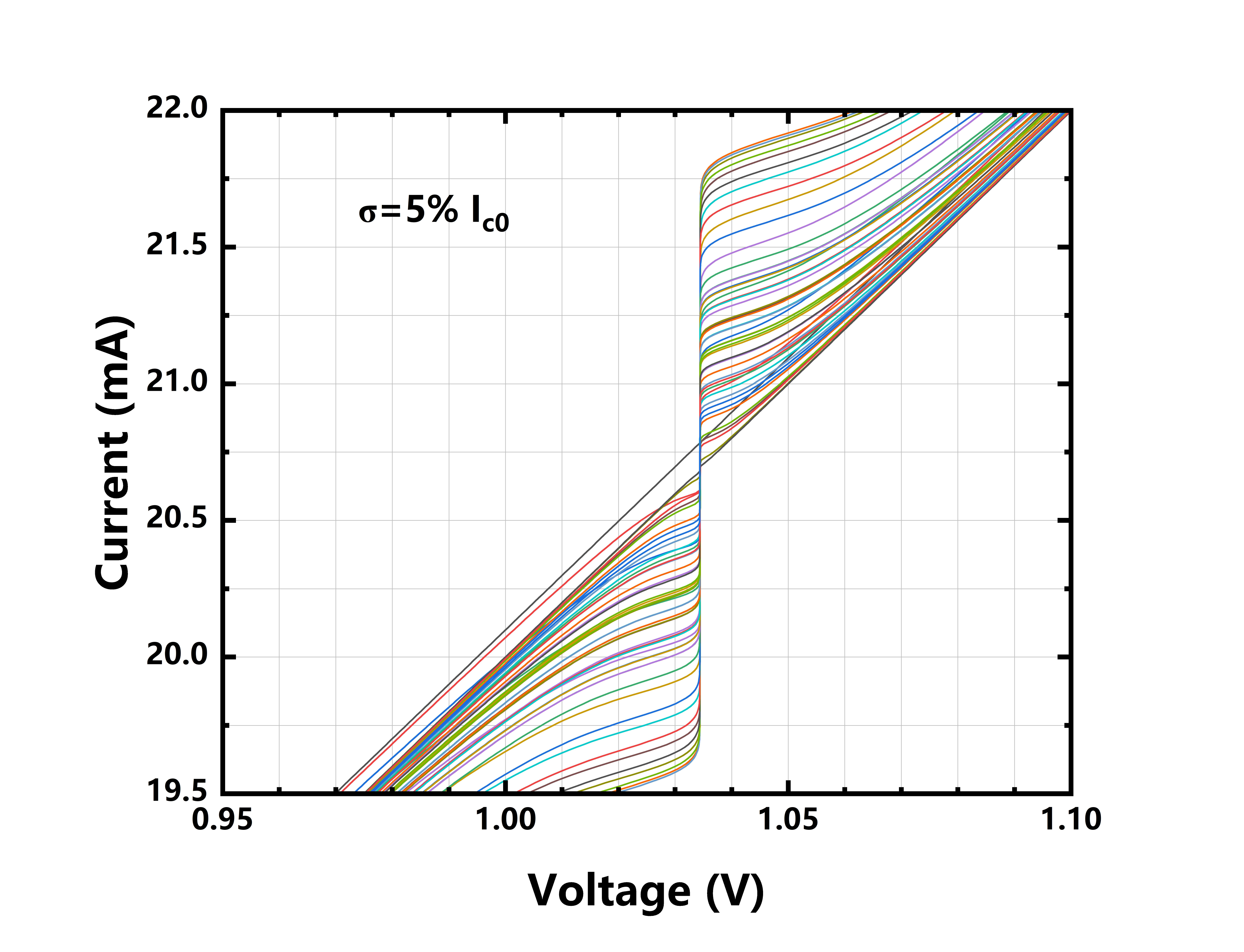}
    \caption{}
    \label{fig4b}
  \end{subfigure}\hfill
  \begin{subfigure}{0.16\textwidth}
    \centering
    \includegraphics[width=\linewidth]{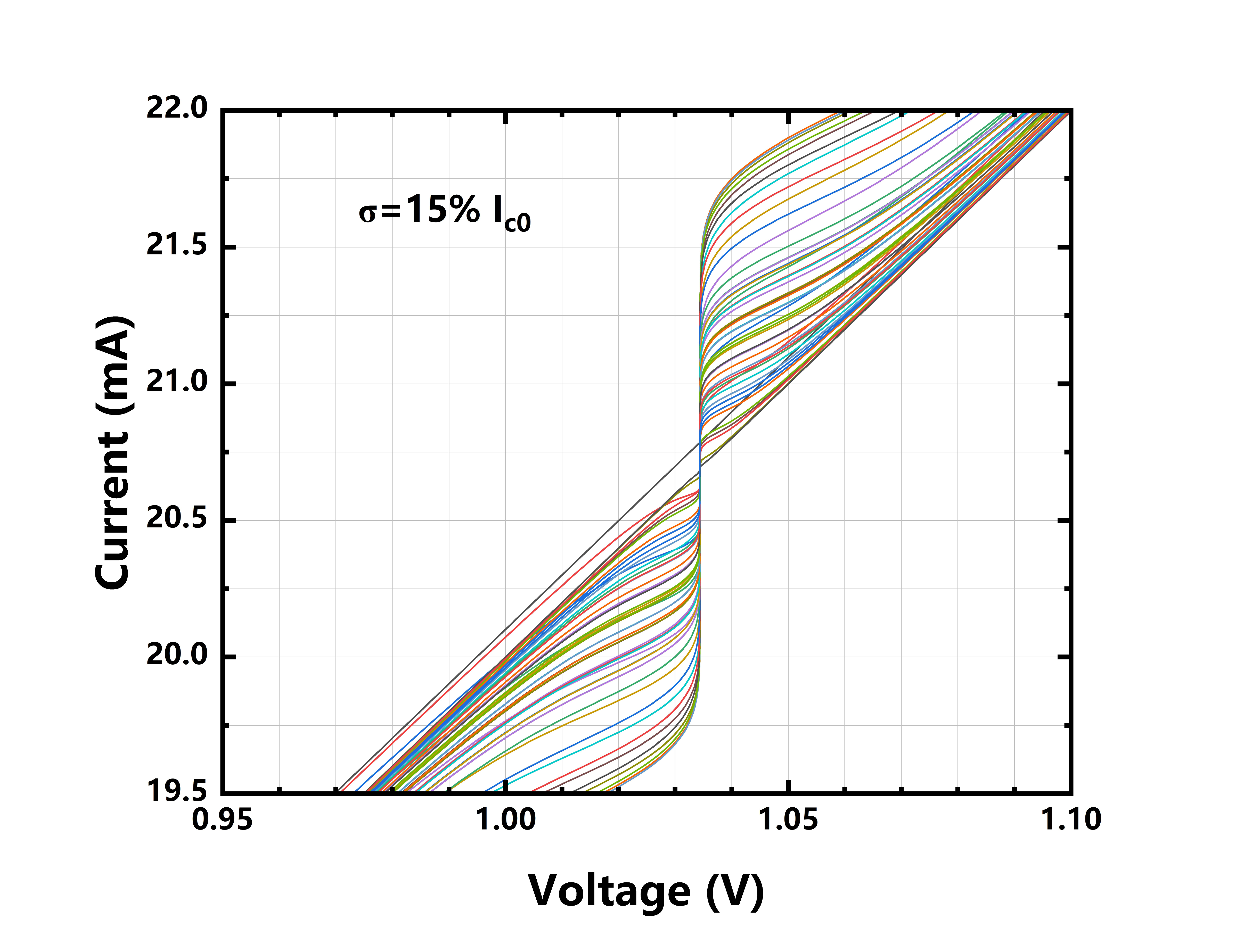}
    \caption{}
    \label{fig4c}
  \end{subfigure}

  \medskip

  \begin{subfigure}{0.16\textwidth}
    \centering
    \includegraphics[width=\linewidth]{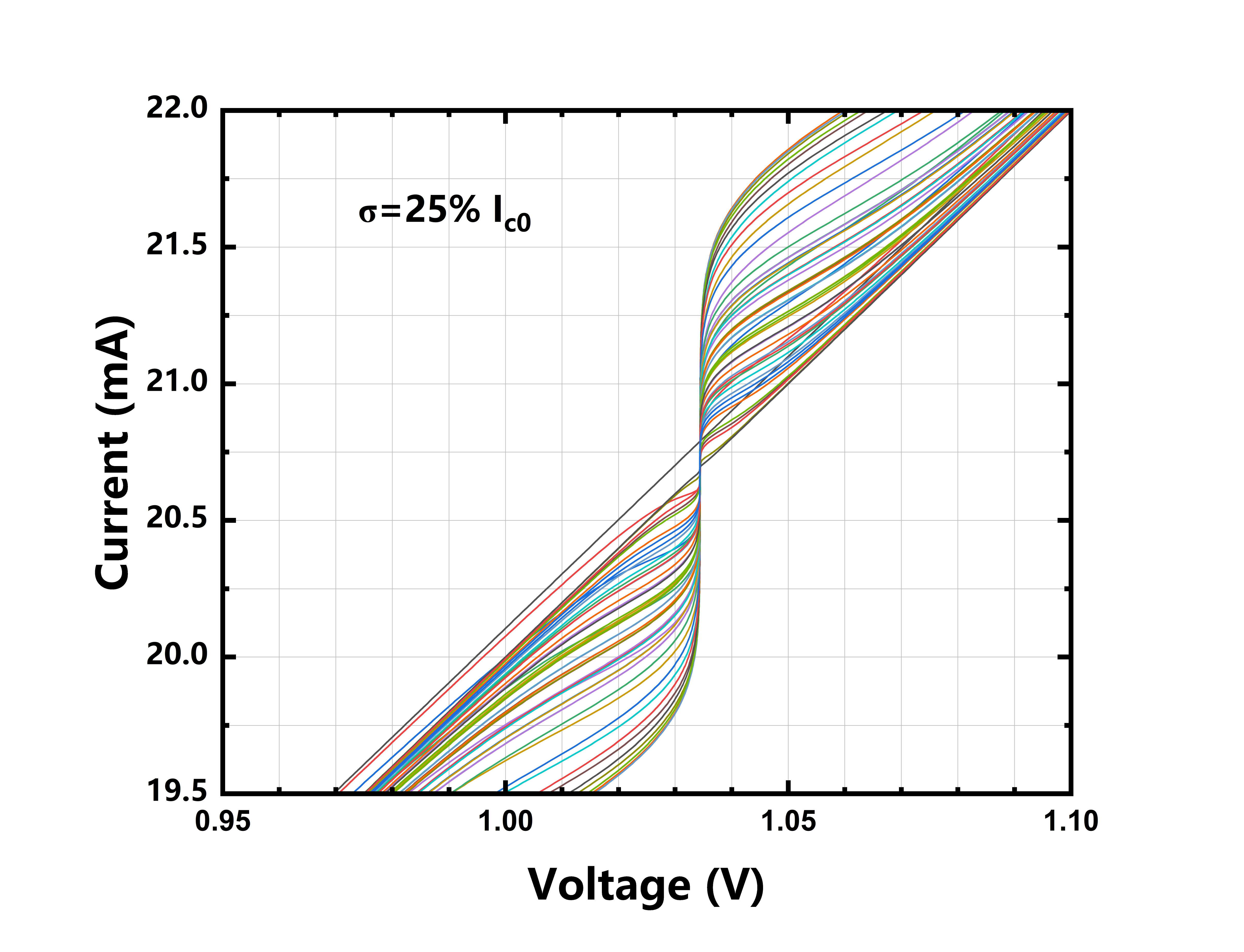}
    \caption{}
    \label{fig4d}
  \end{subfigure}\hfill
  \begin{subfigure}{0.16\textwidth}
    \centering
    \includegraphics[width=\linewidth]{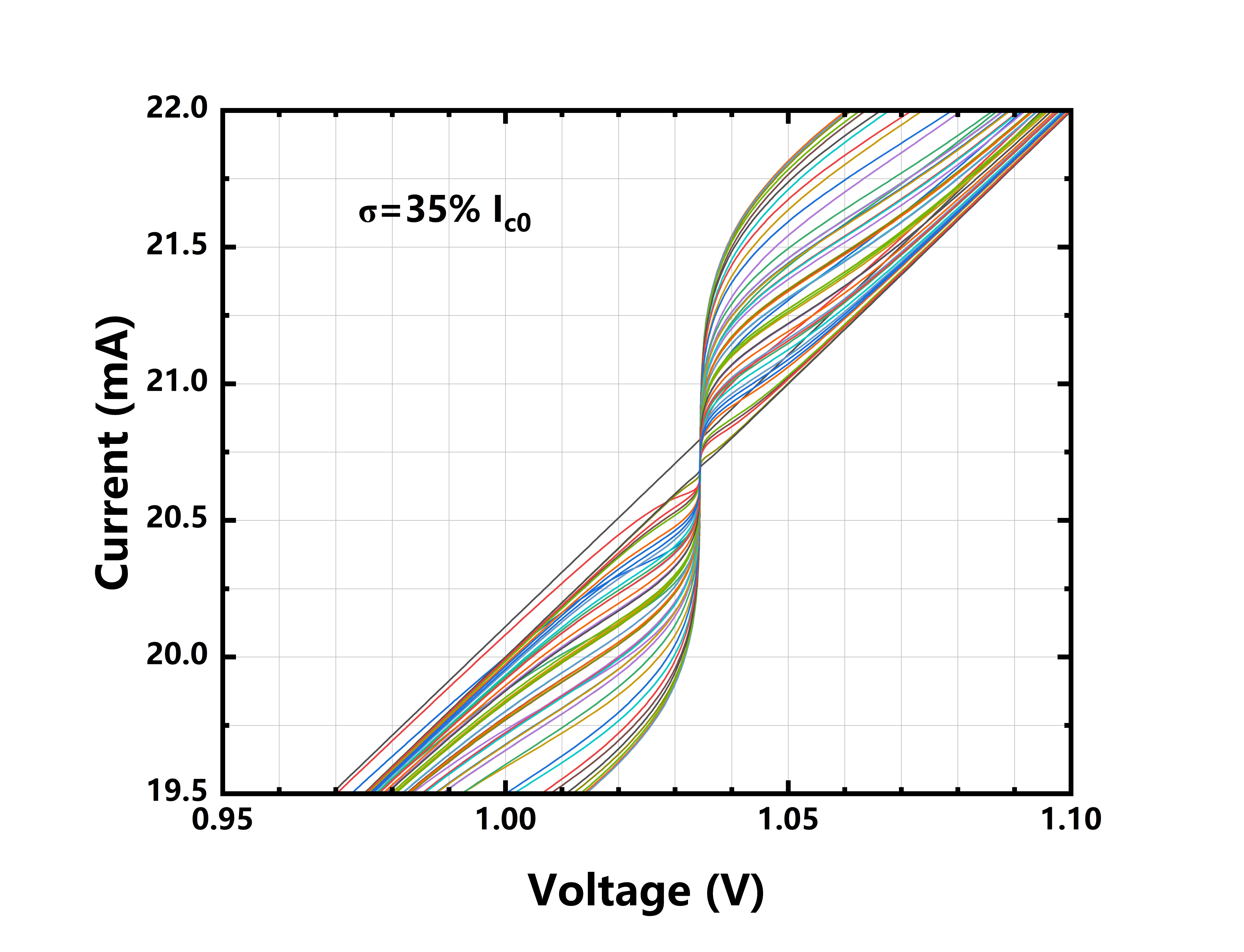}
    \caption{}
    \label{fig4e}
  \end{subfigure}\hfill
    \begin{subfigure}{0.16\textwidth}
    \centering
    \includegraphics[width=\linewidth]{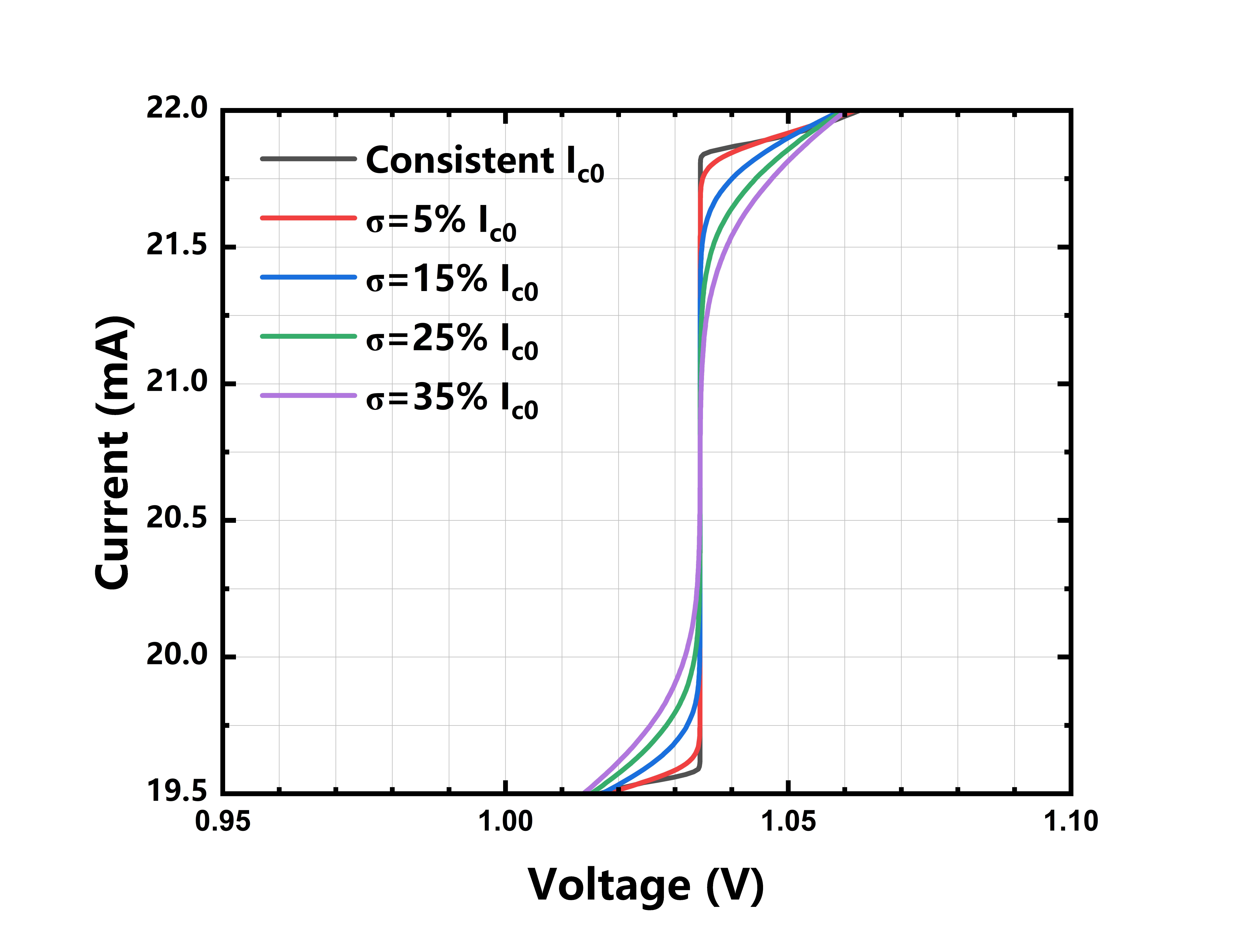}
    \caption{}
    \label{fig4f}
  \end{subfigure}\hfill

\caption{(a-e) $I$-$V$ characteristics with thermal noise at varying $\sigma$ values. The curves in different colors represent distinct microwave amplitude $V_{rf}$. (f) $I$-$V$ characteristic at $\max\Delta I_{1}$ under different $\sigma$.}
\label{fig4}
\end{figure}

\begin{figure}[ht!] 
\centering
\includegraphics[width=2 in]{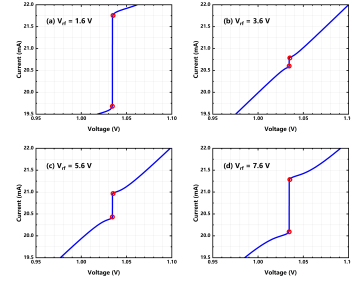}
\caption{The variation of the $\Delta I_{1}$ with the microwave voltage amplitude for different values of $\sigma$ with thermal noise ($I_c$ spread).}
\label{fig5}
\end{figure}

To improve computational efficiency while maintaining accuracy, we concentrated our simulations on the $I$-$V$ characteristics near the first Shapiro step. As shown in Figure \ref{fig4}, the $I$-$V$ characteristics exhibit progressively smoother transitions at the first Shapiro step with increasing $\sigma$ values, accompanied by a correlated decrease in $\Delta I_{1}$ (Figure \ref{fig4f}). According to the formula $V = hf/2e$, the precise position of the first step can be calculated. By selecting a defined range around this position as the interval for determining the step height, we can obtain the step height under the given conditions. As illustrated in Figure \ref{fig5}, the difference in current between the two red circles corresponds to the step height. In this study, the selected range is set to ±10 $\mu$V. It should be noted that a smaller range imposes higher accuracy requirements for the computation of the $I$-$V$ curve. Researchers may adjust this range according to specific research needs and computational capabilities.

\begin{figure}[ht!]
  \centering
  \begin{subfigure}{0.24\textwidth}
    \centering
    \includegraphics[width=\linewidth]{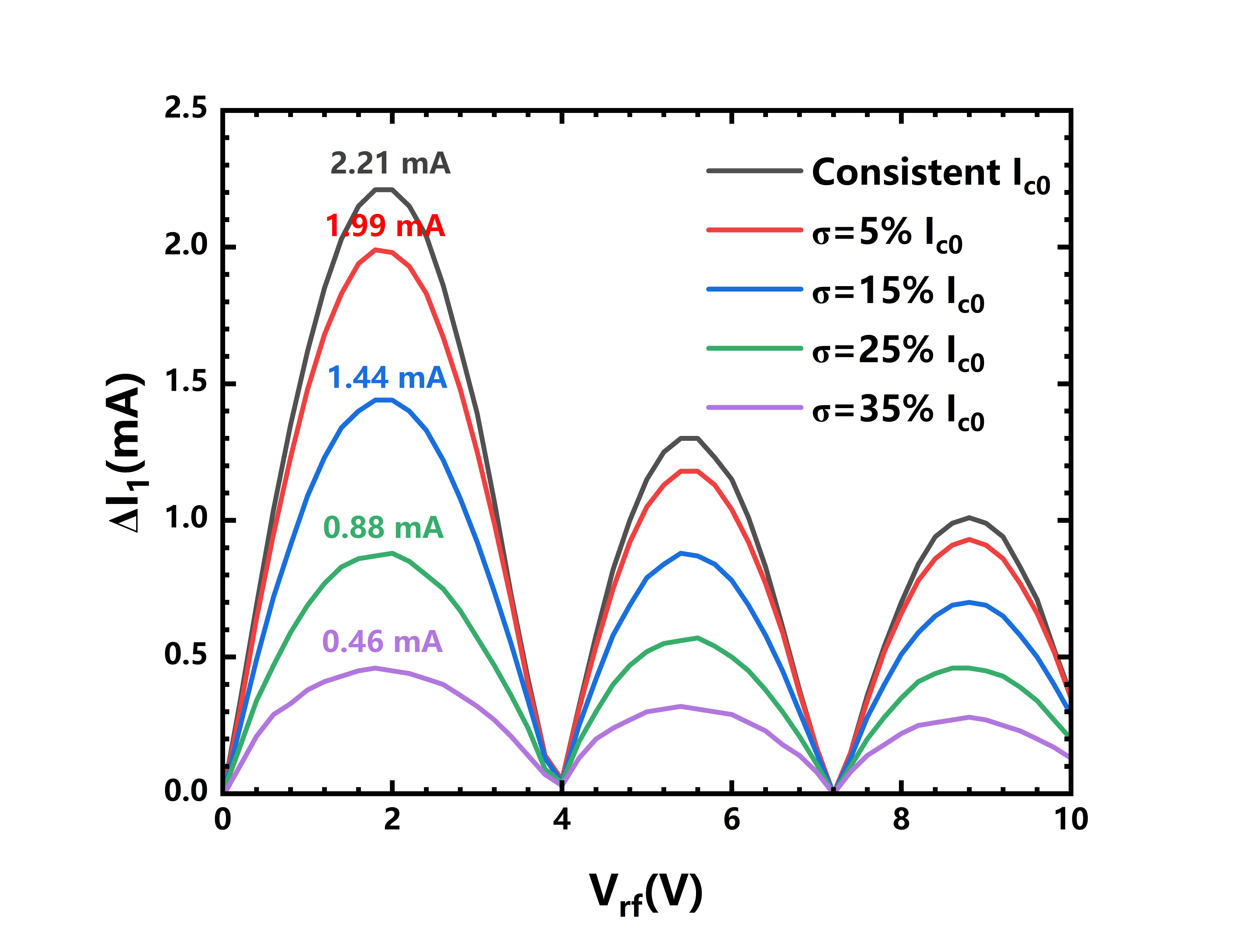}
    \caption{}
    \label{fig6a}
  \end{subfigure}\hfill
  \begin{subfigure}{0.24\textwidth}
    \centering
    \includegraphics[width=\linewidth]{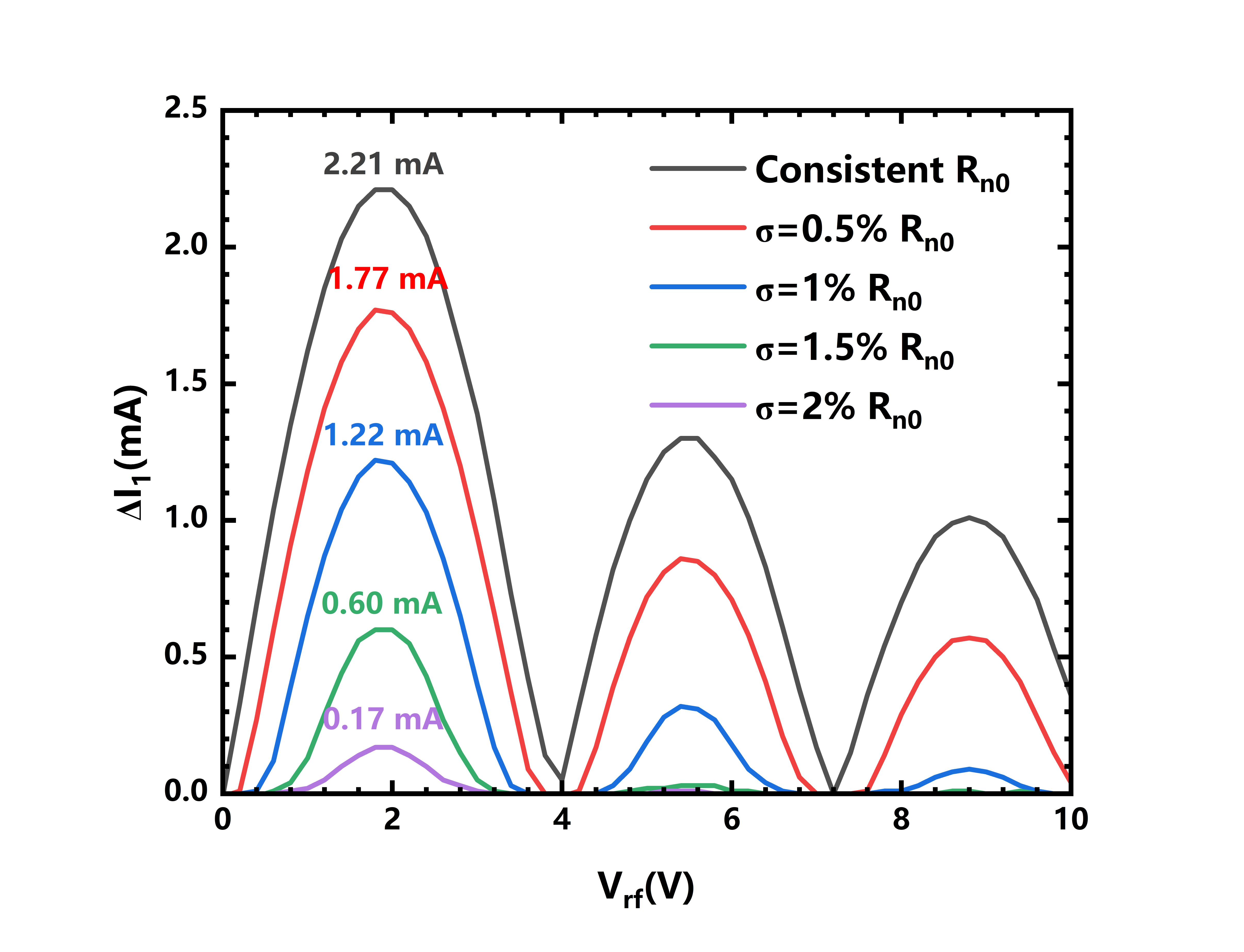}
    \caption{}
    \label{fig6b}
  \end{subfigure}
  \caption{The variation of the $\Delta I_{1}$ with the microwave voltage amplitude for different values of $\sigma$ with thermal noise. (a) $I_c$ spread, (b) $R_n$ spread.}
  \label{fig6}
\end{figure}

Figure \ref{fig6a} systematically examines the dependence of the first Shapiro step height on microwave voltage amplitude for various critical current dispersion conditions. As evidenced by the results, the maximum step height reaches 2.21 mA when the critical current remains uniform. In contrast, with a critical current dispersion of $\sigma$ = 25\%$I_{c0}$, the maximum step height decreases to approximately 0.88 mA. These findings demonstrate that, under the present simulation parameters, maintaining a step height exceeding 0.88 mA requires the critical current distribution to be better than 25\%.

Furthermore, we investigated the influence of the spread in Josephson junction resistance on $\Delta I_{1}$ in the array. The critical current $I_c$ of the Josephson junctions was kept constant at 2 mA, while the mean resistance value was set to $R_{n0}$ = 0.005 $\Omega$, ensuring that the total series resistance remained at 50 $\Omega$. By varying the standard deviation of the resistance distribution, $\sigma$, to 0.5\%, 1\%, 1.5\%, and 2\% of $R_{n0}$, the relationship between the height of the first Shapiro step and the microwave voltage amplitude under different critical current dispersion conditions are presented in Figure \ref{fig6b}. From the figure, it is clear that $\Delta I_{1}$ is significantly affected by the resistance distribution. When $\sigma = 2\% R_{n0}$, the Shapiro first step nearly disappears.


\section{Conclusion}
We quantitatively analyzed the influence of critical current and normal state resistance spread on the Shapiro step height in HTS Josephson junction arrays. The critical current was assumed to follow a normal distribution, $I_c \sim N(I_{c0}, \sigma^2)$, where $I_{c0} = 2 \text{ mA}$ and $\sigma$ took values of 0\%, 5\%, 15\%, 25\%, and 35\% of $I_{c0}$. Numerical simulations were performed on a series-connected array consisting of 10,000 Josephson junctions, with a total resistance of 50 $\Omega$. The simulations were conducted under two conditions: one without noise and the other considering thermal noise at 77 K, with a microwave frequency of 50 GHz.

From the $I-V$ curves obtained for different values of $\sigma$, we observed that as $\sigma$ increased (indicating greater dispersion in the critical current), the transitions in the $I-V$ curve became smoother, and the height of the Shapiro steps decreased. Numerically, when $\sigma$ equaled 0\%, 5\%, 15\%, 25\%, and 35\%, of $I_{c0}$, the corresponding maximum first Shapiro step heights were 2.21 mA, 1.99 mA, 1.44 mA, 0.88 mA, and 0.46 mA, respectively. To achieve a maximum first Shapiro step height of 0.88 mA or higher, the critical current dispersion must not exceed a threshold, specifically, the standard deviation should not exceed approximately 25\% of $I_{c0}$.
Under the conditions of this study, when the resistance distribution had a standard deviation of 0.5\%, 1\%, 1.5\% and 2\% of $R_{n0}$, the maximum height of the first Shapiro step in the Josephson junction array was approximately 1.77 mA, 1.22 mA, 0.6 mA and 0.17 mA, respectively. These results suggest that achieving consistent resistance across the junctions is crucial for the reliable application of JVS.

The results presented in this paper are derived from simulations conducted under specific parameter settings established in this study, such as $I_{c0}$ and $R_{n0}$. More importantly, the primary contribution of this research lies in the development of a novel numerical simulation methodology, which enables detailed studies on the impact of Josephson junction parameter spread on JVS performance. This approach provides a valuable tool for future investigations in this field.




\bibliographystyle{IEEEtran}
\bibliography{IEEEabrv,biblio_traps_dynamics}

\end{document}